\documentclass[a4paper]{article}       % onecolumn (second format)
\usepackage{graphicx}
\usepackage[T1]{fontenc}
\usepackage{amsmath, amssymb, latexsym, booktabs, amsfonts}
\usepackage{url,hyperref,lineno,microtype}
\usepackage[onehalfspacing]{setspace}
\usepackage{array}
\usepackage{xcolor}
\usepackage{authblk}

\newcommand{\idion}{Idiotypic Network }
\newcommand{\goes}[1]{\xrightarrow[]{#1}}
\newcommand{\infr}[2]
	{\renewcommand{\arraystretch}{1.5}
	\begin{array}{c}
	#1\\
	\hline
	#2
	\end{array}}
	
\newcounter{todocounter}

\newtheorem{theorem}{Theorem}[section]
\newtheorem{definition}[theorem]{Definition}

\title{Topological Run-time Monitoring\\ for Complex Systems}

\author[1]{Matteo Rucco\thanks{Corresponding Author: matteo.rucco@utrc.utc.com}}
\author[2]{Luca Tesei\thanks{luca.tesei@unicam.it}}
\author[2]{Emanuela Merelli\thanks{emanuela.merelli@unicam.it}}
\affil[1]{United Technology Research Center, Via Praga 5, 38121 Trento (Tn), Italy}
\affil[2]{School of Science and Technology, University of Camerino\\ Via Madonna delle Carceri 9, 62032 Camerino (MC), Italy}

\date{}

\begin{document}

\maketitle

\begin{abstract}
In this paper we introduce a new data-driven run-time monitoring system for analysing the behaviour of time evolving complex systems. The monitor controls the evolution of the whole system but it is mined from the data produced by its single interacting components. Relevant behavioural changes happening at the component level and that are responsible for global system evolution are captured by the monitor. Topological Data Analysis is used for shaping and analysing the data for mining an automaton mimicking the global system dynamics, the so-called Persistent Entropy Automaton (PEA). A slight augmented PEA, the monitor, can be used to run current or past executions of the system to mine temporal invariants, for instance through statistical reasoning. Such invariants can be formulated as properties of a temporal logic, e.g.\ bounded LTL, that can be run-time model-checked. 
We have performed a feasibility assessment of the PEA and the associated monitoring system by analysing a simulated biological complex system, namely the human immune system. The application of the monitor to simulated traces reveals temporal properties that should be satisfied in order to reach immunization memory.
\end{abstract}

\section{Introduction}

Complex systems are systems made by a potential infinite number of interacting entities, each of them equipped with their own behaviors and strategies. In general, complex systems can not be fully modelled and thus an exhaustive analyses of their behaviors can not be performed.  Some complex systems are classified as {\it self-adaptive systems}, where the ``self'' prefix indicates that the system autonomously decide how to adapt or to re-organize itself so that it can accommodate changes in its contexts and environments.

\begin{figure}[th]
\center
\includegraphics[width=0.5\textwidth]{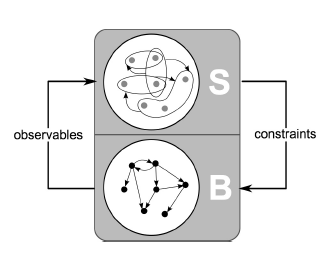}
\caption[S{[}B{]} model]{$S[B]$ model. The picture is extracted from~\cite{merelli2014topology}}
\label{fig:sb}
\end{figure}

In principle, also software systems can be self-adaptive if they are capable, to some extent, of evaluating and changing their own behaviour. Typically, adaptation happens when an evaluation shows that the software is not accomplishing what it was intended to do, or when better functionality or performance may be possible. There are two types of adaptation:
\begin{itemize}
\item structural adaptation, which is related to architectural reconfiguration;
\item behavioral adaptation, which is related to function changes.
\end{itemize}
There are several modeling techniques for complex systems, for a complete overview we refer to \cite{boccara2010modeling}. In~\cite{merelli2012multi,MPT15}, two of the authors have introduced a new modelling paradigm, the so-called $S[B]$ paradigm,  for modelling self-adaptive systems. In the $S[B]$ paradigm a model has two levels of description, namely the $S$ \emph{global} or \emph{structural} level and the $B$ \emph{local} or \emph{behavioral} level, which are entangled in order to express the behavior of the system as a whole. The $S$ level describes how the system evolves following global information coming from the environment in which it is operating and from the interactions and the evolutions of the entities of the $B$ level. A graphical description of an $S[B]$ model is given in Figure~\ref{fig:sb}.

%%%%%%%%%%%%%%
%%%%%%%%%%%%%%

In the $S[B]$ paradigm, the $B$-level adapts itself according to the higher level rules. The $S$-level constraints the $B$-level. On the behavioral level, adaptation is expressed by firing a higher-order transition, meaning that the $S$-level switches to a different set of constraints and the $B$-level has adapted its behavior by reaching a state that meets the new constraints. This adaptation is not necessarily instantaneous. Moreover, the $S$-level can exhibit adaptation by switching from a model to another or by redefining the set of invariants. In~\cite{MPT15} the authors proposed the first formal definition of $S[B]$ by means of finite state machines both for  the behavioural level and the structural level. 

In general, any model of a complex system can be subsequently enriched by information extracted from observations. Typically, the process of extracting additional information is referred to as \emph{specification mining} \cite{reger2013pattern}. Specification mining examines the execution traces of a system to determine patterns of event occurrences. These motifs are called \textit{specifications} and identify some aspects of the behaviour of the system. The examination of traces can occur at run-time or through post-run event logs and  the process of specification mining should be performed throughout the entire life of a system.  

When complex self-adaptive systems are used in safety critical contexts (e.g., therapy administration, autonomous-driving and so on) it is not possible to wait until the end of their execution and, thus, there is the need to equip them with run-time monitoring procedures that produce verdicts about the violation of safety properties. With the increasing of the complexity of the systems, new run-time monitoring techniques must be developed for demonstrating the reliability of these systems w.r.t.\ safety assessment standards (e.g. ISO26262, ARP4761) as requested by certification bodies \cite{koopman2018toward}. From a verification point of view, adaptability can be seen as a particular case of run-time verification. Run-time verification is obtained by equipping a system with a monitor that recognizes when the system violates certain properties. If this happens the monitor possibly sends a feedback to the system in order to recover the attended state~\cite{de2013software}. 

In this work, based on $S[B]$, we define and apply a new data driven run-time monitoring system for controlling time evolving complex systems. The monitor controls the evolution of the entire system but it is mined from the data produced by the single interacting components that shape the whole system. Relevant behavioural changes happening at the component level and that are responsible for global system evolution are captured by the monitor. Topological Data Analysis (TDA) is used for shaping and analysing the data. TDA finds topological invariants, i.e., persistent barcodes and Persistent Entropy (PE), which are used for defining an automaton mimicking the system dynamics, the so-called Persistent Entropy Automaton (PEA). From the PEA we show how to derive the monitor system that can run current or past execution of the system. From the analysis of the resulting traces, e.g.\ through statistical reasoning, it is possible to mine temporal invariants. The invariants can then be formalized by means of a temporal logic suitable for run-time verification. 

For the sake of completeness, this work summarizes the main finding of the PhD thesis of M. Rucco \cite{matteorucco2016}. The paper is organized as follows. In Section~\ref{sec:background} we introduce the relevant background, namely the fundamental concepts of TDA and the structure of run-time verification. Section~\ref{sec:ImmSys} introduces the case study of the human Immune System and the \idion simulator that we use to produce the analysed dataset. The PEA and its formal semantics is then defined in Section~\ref{sec:pea} together with the recall of the derivation of the PEA for the case study already introduced in \cite{merelli2015topological}. The PEA is used to define a monitor system in Section~\ref{sec:monitor} where the \idion case study is used to reveals temporal properties that should be satisfied in order to reach a condition that is the immunization memory. 

\section{Background}
\label{sec:background}

\subsection{Topological Data Analysis}
A topological space is a powerful mathematical concept for describing the connectivity of a space. Informally, \textit{a topological space} is a set of points, each of which equipped with the notion of \textit{neighboring}. 
Algebraic topology uses typical concepts of abstract algebra to study topological spaces~\cite{hatcher2002algebraic, munkres1984elements}. In the last decade a new suite of tools, based on algebraic topology, for data exploration and modeling haven been invented~\cite{carlsson,zomorodian2007topological,edelsbrunner2008persistent}. The data science community refers to these tools as \textit{Topological Data Analysis} (TDA). TDA has been used in different domains: biology, manufacturing, medicine and others~\cite{rucco2016survey}. In the following subsections we recall the mathematical definitions of the main concepts needed for this paper.

\subsubsection*{Persistent homology}
Homology is an algebraic machinery used for describing a topological space $\mathfrak{C}$.
%The $k-$Betti number represents the rank of the $k-$dimensional homology group.
Informally, for a fixed natural number $k$,
the $k-$\textit{Betti number} $\beta _k$ counts
 the number of $k-$dimensional holes characterizing $\mathfrak{C}$:
$\beta_0$ is the number of connected components, $\beta_1$ counts the number of
holes in 2D or tunnels in 3D\footnote{Here $n$D refers to the $n-$dimensional space $\mathbb{R}^n$.}, $\beta_2$ can be thought as the number of voids in geometric solids, and so on.

Persistent homology is a method for computing the $k-$dimensional holes at different spatial resolutions. Persistent holes are more likely to represent true features of the underlying space, rather than artifacts of sampling (noise), or due to particular choices of parameters.
For a more formal description we refer the reader to~\cite{edelsbrunner2010computational}. In order to compute persistent homology, we need a distance function on the underlying space. This can be obtained constructing \textit{a filtration} on a simplicial complex, which is a nested sequence of increasing subcomplexes. More formally, a filtered simplicial complex $K$ is a collection of subcomplexes
$\{K(t): t \in \mathbb{R}\}$ of $K$ such that $K(t) \subset K(s)$ for $t< s$ and there exists $t_{max}\in \mathbb{R}$ such that $K_{t_{max}}=K$. The filtration time (or filter value) of a simplex $\sigma \in K$ is the smallest $t$ such that $\sigma \in K(t)$.

Persistent homology describes how the homology of $K$ changes along a filtration. A $k-$dimensional Betti interval, with endpoints $[t_{start}, t_{end}),$ corresponds to a $k-$dimensional hole that appears at filtration time $t_{start}$ and remains until time $t_{end}$. We refer to the holes that are still present at $t= t_{max}$ as \textit{persistent topological features}, otherwise
they are considered \textit{topological noise}~\cite{adams2011javaplex}.
The set of intervals representing birth and death times of homology classes is called the {\it persistence barcode} associated to the corresponding filtration. 
Instead of bars, we sometimes draw points in the plane such that a point $(x,y)\in \mathbb{R}^2$ (with $x< y$) corresponds to a bar $[x, y)$ in the barcode. This set of points is called {\it persistence diagram}. There are several algorithms for computing persistent barcodes, the principal of which are Gudhi~\cite{maria2014gudhi} and jHoles~\cite{binchi2014jholes}. For a complete overview of the available tools we refer to~\cite{otter2017roadmap}.

\subsubsection*{Persistent Entropy}
In order to measure how much the construction of a filtered simplicial complex is ordered a new entropy measure, called \textit{Persistent Entropy} (PE), was defined in~\cite{rucco2015characterisation}. A precursor of this definition was given in~\cite{chintakunta2015entropy} to measure how different bars of a barcode are in length. Here we recall the definition.

\begin{definition}[Persistent Entropy] \ \\
Given a filtered simplicial complex $\{K(t) :  t\in F\}$, and the corresponding persistence barcode $B = \{a_i=[x_i , y_i) : i\in  I\}$, the \textit{Persistent Entropy} (PE) $H$ of the filtered simplicial complex is defined as follows:
$$
H=-\sum_{i \in I} p_i  log(p_i)
$$
where $p_i=\frac{\ell_i}{L}$, $\ell_i=y_i - x_i$, and $L=\sum_{i\in I}\ell_i$.
\end{definition}
Note that, when topological noise is present, for each dimension of the persistence barcode there can be more than one interval, denoted by $[x_i~,~y_i)$ with $i \in I$. This is equivalent to say that, in the persistence diagram, the point $[x_i,y_i)$ could have multiplicity greater than $1$ (see \cite[page 152]{edelsbrunner2010computational}). 

In the case of an interval with no death time, $[x_i,+\infty)$, 
several approaches can be considered, such as extending real numbers including $+\infty$,  removing or truncating infinite intervals or using extended persistence \cite{extending,chazal,pawel}. In this paper, we  truncate infinite intervals and replace $[x_i,+\infty)$ by $[x_i~,~m)$ in the persistence barcode, where $m = t_\mathrm{max} + 1$.

Note that the maximum PE corresponds to the situation in which all the intervals in the barcode are of equal length. In that case, $H=\log n$ if $n$ is the number of elements of $I$. Conversely, the value of the PE decreases as more intervals of different length are present.

The stability theorem defined in~\cite{rucco2017new} defines how to compare persistent entropies from different simplicial complexes. The main requirement is that the simplicial complexes have the same number of 0-simplices (nodes). PE is a  tool, equipped with suitable mathematical properties, that permits to describe complex systems~\cite{atienza2019persistent} and it has been applied in different experiments, e.g.\ the analysis of biological images~\cite{jimenez2017topological} and the analysis of medical signals~\cite{piangerelli2018topological}.

\subsection{Run-time Verification}
\label{RTV}
{\it Model checking} and {\it run-time verification} are two different methods for performing formal analysis on systems. Model checking is executed statically on a given formal model of the systems under study. The verification task takes a formally specified property and exhaustively checks whether or not all the possible behaviors associated to the model satisfy the property. If not, a counterexample trace is given by the model checking algorithm as output which shows an admissible behavior of the model that leads to the violation of the property. Usually, temporal logics are used for specifying properties and labeled transition systems are the basic semantic structures on which the model is defined. For a more general introduction on model checking we refer to~\cite{baier2008principles}. 

In contrast to model checking, run-time verification examines a system dynamically and extracts information from its executions to detect violations. It inherently explores only a portion of the state space, thus its overall guarantees are weaker than those offered by model checking. The properties that are checked need only to hold over the portion of the state space that is visited during the execution of the system once it is deployed. The class of properties amenable to run-time verification is a strict subset of the properties amenable to model checking. For example, liveness properties, i.e.\ those requiring that ``something good will eventually happen'', cannot properly be decided on a finite execution trace. Thus, only \emph{bounded} liveness properties can be checked at run-time, i.e.\ those requiring that ``something good will happen within a specified number of steps''. On the other hand, for safety properties, i.e.\ those requiring ``nothing bad happens'', a violation can be decided on a finite trace. 

In run-time verification, a property $\psi$ is translated into a monitor, i.e.\ an automaton that is put in parallel with the system during its execution and observes the system moves. From a {\it formal language} viewpoint, run-time verification is similar to to the {\it word problem}, that is deciding whether or not a given word is included in a given language. More precisely, given the set $W$ of finite executions of the model, run-time verification checks if an execution $w \in W$ is an element of $\mathbb{L}(\psi)$, i.e.\ the language of all finite executions satisfying the property $\psi$. 

%%%%%%%%%%%%%%%%%%
\begin{figure}[ht!]
\begin{center}
\includegraphics[scale=.5]{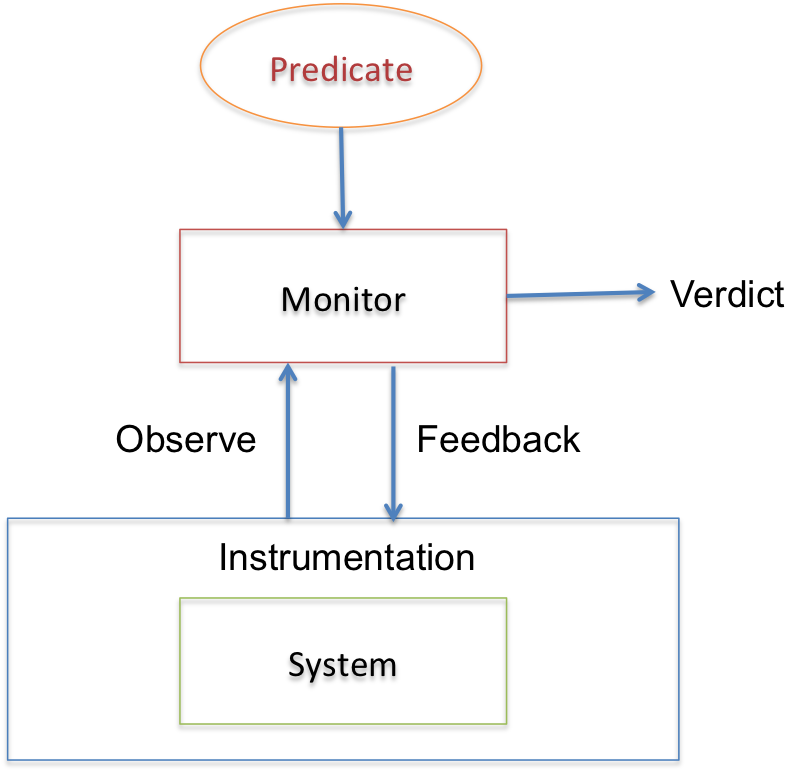}
\caption[run-time process]{Overview of the run-time verification process.}
\label{fig:rvclassic}
\end{center}
\end{figure}
%%%%%%%%%%%%%%%%%%

A run-time verification process typically consists of the following steps (see Figure~\ref{fig:rvclassic}):
\begin{enumerate}
\item Monitor synthesis: from a logical property a monitor is generated. The monitor is capable of consuming {\it events} produced by a running system and emits verdicts according to the current satisfaction of the property.
\item System instrumentation: the aim of this stage is to be able to generate the relevant events to be fed to the monitor.
\item Execution analysis: the execution of the system is analyzed by the monitor. When the monitor is used to check $\psi$ on the system during its execution, in an incremental fashion, we speak of {\it on-line run-time verification}. Instead, a monitor can also be used to check a {\it finite} set of recorded executions w.r.t.\	 $\psi$, which is known as {\it off-line run-time verification}. The monitoring can be attended by using {\it logging systems}, {\it trace tools}, or {\it dedicated tracing hardware}. 
\end{enumerate}
For a more detailed introduction to run-time verification we refer the reader to~\cite{delgado2004,leucker2009brief,leucker2012teaching,cassar2017asurvey,bartocci2018introduction}.

\section{Case Study: Human Immune System}
\label{sec:ImmSys}
The precise function of the human Immune System (IS) remains undetermined; it is postulated that it plays at least two roles: it protects against invading micro-organism and it regulates bodily functions. Recently, scientists have highlighted also function interconnection between the IS and the brain activities. Previously, IS was described as a system based on two separated phases, respectively known as {\it innate response} and {\it adaptive response}, but recent discoveries advocated that thanks to the interaction between these two responses the whole system can maturate the high level protection required for the maintenance of the healthy status~\cite{perelson1989immune,perelson1997immunology,hoffmann1975theory,rapin2011immune,mancini2012,castiglione2015}. 

In the \idion, the concept of liveness that was introduced in Sec. \ref{RTV} corresponds to the ability of the system to force itself in order to reach the {\it memory state}. 

Beside the biological features, the IS shows all the features characterizing a complex system: {\it robustness, concurrency, decentralization, fault tolerance,  adaptability, and multiple roles}. 
There exist several mathematical (e.g., differential equation, mean-field, network) and computational (e.g., cellular automata) models of the IS. In this work we use {\it C-ImmSim}, which is a multi-agent system for simulating the IS\footnote{The authors kindly acknwoledge Dr. Filippo Castiglione.} and is able to simulate all the features highlighted above.

\subsection{Immune Network Theory}

Lindenmann and Jerne proposed a theory describing the immune system as a network of interaction of antibodies and lymphocytes. According to this hypothesis, Idiotype and anti-Idiotype network interactions would regulate the immune response of the host against a given antigen.

\begin{figure}[htpb!]
  \centering
  \includegraphics[width=\textwidth]{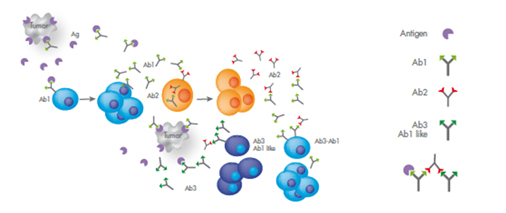}
  \caption[Humoral Immune Response]{Example of Idiotypic Network~\cite{vaxira}.}
  \label{fig:immresponse}
\end{figure}

In the \idion model the immunization for a given antigen Ag will generate the production of antibodies, called Ab1, against this antigen (see Figure~\ref{fig:immresponse}). This Ab1 can generate a series of anti-Id antibodies against Ab1, called Ab2. The particular anti-Ids which fit into the antigen binding site of Ab1 can induce a specific immune responses against the nominal antigen Ag. A practical consequence is that these idiotopes could be used to mimic any existing antigen and used as surrogate antigens. Immunization with Ab2 can lead to the generation of anti-anti-Id antibodies (called Ab3) that recognize the corresponding original antigen identified by Ab1. For instance, several such Ab2 have been used to trigger the immune system to induce specific and protective immunity against tumor antigens. The above description has been formalized with different approaches~\cite{parisi1990simple,merelli2014topology,wherry2003lineage,fachada2007agent}. 

\subsection{C-ImmSim}
\label{sec:CimmSimm}
C-ImmSim is an agent-based simulator of the mammalian immune system that extends and improves {\it ImmSim} and it is nowadays the {\it de-facto} agent-based model used by immunologists~\cite{rapin2011immune,mancini2012,castiglione2015}. {\it C-ImmSim} implements several models of the IS: (e.g., \idion theory, Clonal selection theory, T cell anergy, and so on) and is highly configurable. 
In {\it C-ImmSim} the entities, at cellular level, are represented by agents, i.e., they keep their individual identity throughout their simulated life span. Cells interact locally in a {\it 3D-lattice}. The simulator allows simulations managing millions of cells. Lymphocytes T helper, TH, cytotoxic, CTL, lymphocytes, antibody-producing plasma cells, PLB, macrophages, MA, dendritic cells and DC are considered, with a very high degree of complexity. Each cell is represented by a bit-string of length {\it NBITS}, so the potential repertoire size is $2^\mathit{NBITS}$. From a computational perspective, {\it C-ImmSim} w.r.t.\ ImmSim, is characterized by an improved efficiency, simulation size and complexity. {\it C-ImmSim} models and detects the presence of antigens, by activating the humoral and cellular responses. 
{\it C-ImmSim} implements the following dynamics of a polyclonal model: the lymphocytes are created in the bone marrow compartment and their receptor is randomly chosen in the bit string space. The B lymphocytes, after the stimulation, go into circulation directly, while the T lymphocytes pass through the Thymus where they undergo selection for auto reactive cells. The mobility of cells is modeled by taking into account realistic diffusion coefficients observed in vivo. T cells have a faster diffusion constant than B cells. Moreover, in absence of chemotaxis, macrophage and dendritic cells have diffusivity similar to B cells. The time scale is obtained by mapping a tick to few hours of real life. However, the user can choose to have 8, 4, 2 or 1 hour as a definition of a simulated time step. Thanks to the bit-string representation, the {\bf affinity} is defined as follows. Two bit-strings complement each other (or are a perfect match) if every 0 in one corresponds to a 1 in the other and conversely. More generally, an $m$-bit match is defined as a pair where exactly $m$ bits complement each other. The affinity is then defined as a monotonic function of $m$; the higher is $m$ the more is the likelihood for the two strings to bind. If the similarity is less than a threshold value, no link occurs. This threshold is indicated as {\bf min-match}. The number of cells per type per step of simulation is implemented by homeostatis. Homeostatis is defined stochastically by taking into account the individual half-life of each cell.
In {\it C-ImmSim} the {\bf immune memory} is implemented as a cell's state acquired during active participation to successive (and successful) immune responses. The effects of the immune memory are directly devoted to increase the half-life of a cell by adding a certain amount every time the cells participate in a successful interaction. The rationale behind this modeling choice is that useful cells survive longer than useless ones simply because they get a whole lot of stimulations during the immune reactions. 

\section{Persistent Entropy Automaton}
\label{sec:pea}

In~\cite{merelli2015topological} the authors introduced a methodology for the data-driven derivation of a Persistent Entropy Automaton (PEA) from a data collection evolving over time. The methodology consists of the following steps:
\begin{enumerate}
\item identify and collect a stream of data produced by the complex system under study;
\item represent the time-dependent dataset - regardless its dimension - as a weighted network (e.g., by computing similarities, correlations, and so on) among the points;
\item compute persistent homology via the weighted rank clique homology algorithm;
\item compute the PE;
\item repeat the previous steps for each observation over time;
\item plot the PE values versus time;
\item identify the segments where the first derivative of the PE is equal to zero. These segments correspond to steady states of the system. There might exist multiple steady states, each of which characterized by their own value of PE.
\end{enumerate}
In general, two steady states with different PE values correspond to two different steady states of the system and are represented by two states of the PEA connected by a transition. Two subsequent steady states with the same value of the PE but separated by a segment with a non zero first derivative of the PE are represented by the same state equipped with a self-loop.

In this work we complete the definition of the PEA given in~\cite{merelli2015topological} with the precise definition of the semantics of the PEA evolution. Moreover, we update the PEA definition to make it slightly more general. In particular, we permit to define steady states not only by recognising the plateaus in the PE plot, but through the specification of general \emph{equilibrium conditions} based on the value of the PE and of its first derivative.

In order to formalize the PEA, its semantics, and, later, the run-time verification based on it, we need to introduce a proper concept of \emph{trace}. In the field of formal models for software/hardware systems, an action-based setting could be adopted in which a trace of the automaton or of the transition system modeling the real system corresponds to the sequence of actions that labels the transitions of a certain path. If a state-based setting is adopted, instead, a trace is usually taken as the sequence of labelling associated to the states of a certain path, typycally a set $AP$ of boolean atomic propositions that are true in the state \cite{de1990action}. In our setting, the steady states and the transitions of the model are derived from the dataset and do not naturally carry associated information on what the modelled system ``does'' or ``satisfies''. Instead, the behaviour of the system is derived from the values of certain \emph{observables} that can be extracted from the data that the system produced. Thus, in general, we assume that a trace is just a sequence of such observables. 

We use $\mathbb{O}$ to denote a generic \emph{observable space}, whose content is determined by the particular complex system that is analysed. 

\begin{definition}[Traces in an Observable Space] \ \\
\label{def:traceObs}
A \emph{trace} $tr$ in an observable space $\mathbb{O}$ is a finite or infinite sequence
$$
tr = o_1 \goes{} o_2 \goes{} \cdots \goes{} o_k \goes{} \cdots
$$
where $o_i$ are observations.
\end{definition}

In our methodology, we analyse the dataset to extract only one observable, namely the PE. Thus, in this paper, the observable space will consist of just time-stamped PE values. 

\begin{definition}[Persistent Entropy Traces] \ \\
\label{def:PEAObs}
A \emph{Persistent Entropy Trace} (PET) is a trace in which each observable is of the form $o=(t,H(t))$, where:
\begin{itemize}
\item[-] $t$ is a time value in a time domain $\mathbb{T}$;
\item[-] $H(t)$ is the value of PE calculated at time $t$;
\end{itemize}
\end{definition}
Thus, a PET is a sequence of the form
$$
tr = (t_1,H(t_1)) \goes{} (t_2,H(t_2)) \goes{} \cdots \goes{} (t_k,H(t_k)) \goes{} \cdots
$$
Note that from the sequence of values of the PE it is also possible to derive a discrete estimation of the first derivative of $H(t)$ by using finite differences.

\begin{definition}[Persistent Entropy Automaton] \ \\
A \emph{Persistent Entropy Automaton} (PEA) is a tuple $(R, \Lambda, \rho_0, \goes{}_S, L)$ where:
\begin{itemize}
\item $R$ is a set of steady states;
\item $\Lambda$ is a set of names for the transitions;
\item $\rho_0 \in R$ is the initial steady state;
\item $\goes{}_S \subseteq R \times \Lambda \times R$ is a labelled transition relation among steady states;
\item $L(\rho)$ is a labeling function associating to each state $\rho \in R$ its \emph{equilibrium condition}, depending on the values of the PE.
\end{itemize}
\end{definition}
In a PEA, a notion of time is intrinsically present, which in general can be continuous or discrete. A value $H$ of the PE must be defined for each instant $t$ in the time domain $\mathbb{T}$. In our methodology \emph{the time domain is discrete} and each discrete instant $t$ corresponds to an observation of the data, i.e.\ the value of the PE is the one calculated from persistent homology at each iteration. The \emph{equilibrium conditions} can be any boolean combination of predicates defined on $H$ and on $\dot{H}$, the first derivative of $H$ w.r.t.\ time. The value of $\dot{H}$ is approximated as the difference quotient between the last two observations of $H$. 

The dynamics of a PEA is informally described as follows. Whenever a PEA is in a steady state $\rho$, the current values of $H$ and $\dot{H}$ must satisfy the associated equilibrium condition $L(\rho)$. Time can elapse while the PEA stays in $\rho$ and the values of $H$ (and, thus, of $\dot{H}$) change accordingly. If at a certain point the equilibrium condition $L(\rho)$ is not satisfied anymore, the PEA must exit the state $\rho$ and start executing a transition $\rho \goes{}_S$ towards one of the possible successor states of $\rho$. This process is in general \emph{non-instantaneous}: the PEA needs some time, i.e.\ further changes in the value of $H$, in order to reach a point in which the equilibrium condition $L(\rho')$ of some successor state $\rho'$ is satisfied. At that moment, the PEA is again in a steady state and the dynamics is again the one previously introduced. More formally, we define the semantics of a PEA by an associated labelled transition system. 

	\begin{definition}[Persistent Entropy Labelled Transition System] \ \\
	\label{def:peasemantics}
A Persistent Entropy Labelled Transition System (PELTS) is a labelled transition system $(Q,\rightarrow, \Sigma)$ where:

		\begin{itemize}

\item $Q = \mathbb{T} \times R \times \mathbb{B} \times \mathbb{R}_{[0,1]}$ is the set of states, where $R$ is a set of states (or locations, to avoid overloading the word ``state'') and $\mathbb{B} = \{\mathit{true},\mathit{false}\}$ is the boolean domain;
\item $\Sigma = \mathbb{T} \times \mathbb{R}_{[0,1]} \times (\Lambda \cup \{\epsilon\})$ is the set of labels, where $\Lambda$ is a set of names and $\epsilon$ is the empty name; and 
\item $\rightarrow \subseteq Q \times \Sigma \times Q$ is a labelled transition relation associating a state to some successor states. 
	
		\end{itemize}

In a state $q= (t,\rho, b , h) \in Q$:
	
		\begin{itemize}
	
\item $t$ is the current timestamp;
\item $\rho$ is a PEA location representing the current steady state or the last visited steady state (when executing a transition);
\item $b$ is a boolean value indicating whether or not the associated PEA is currently in a steady state ($\mathit{true}$) or is currently performing a transition ($\mathit{false}$); and
\item $h$ is the last observed value of PE.
		
		\end{itemize}

	\end{definition}

We associate to a given PEA a PELTS, defining its semantics, as follows.

	\begin{definition}[PELTS associated to a PEA] \ \\
	
Let $(R, \Lambda, \rho_0, \goes{}_S, L)$ be a PEA. The associated PELTS is the labelled transition system $(\mathbb{T} \times R \times \mathbb{B} \times \mathbb{R}_{[0,1]}, \rightarrow, \mathbb{T} \times \mathbb{R}_{[0,1]} \times (\Lambda \cup \{\epsilon\}))$ where $\rightarrow$ is defined as the minimum relation satisfying the rules in Table~\ref{table:rules-PELTS}. 

The PELTS state $q_0= (0, \rho_0, \mathit{true}, 0)$ is called \emph{initial}.

	\end{definition}
	
	\begin{table}
$$
\begin{array}{|c|}
\hline
\mathrm{(Steady)} \, \infr{\begin{array}{c} t < t' \quad \dot{h} = (h' - h) / (t' - t) \\ h', \dot{h} \models L(\rho)\end{array}}
{(t,\rho,\mathit{true},h) \goes{(t',h',\epsilon)} (t',\rho,\mathit{true},h')} \\ 
\hline
\mathrm{(StartT)} \, \infr{\begin{array}{c} t < t' \quad \dot{h} = (h' - h) / (t' - t) \\ h', \dot{h} \not \models L(\rho) \quad (\rho, \lambda, \rho') \in \goes{}_S \end{array}}
{(t,\rho,\mathit{true},h) \goes{(t',h',\epsilon)} (t',\rho,\mathit{false},h')} \\
\hline
\mathrm{(ContT)} \, \infr{\begin{array}{c} t < t' \quad \dot{h} = (h' - h) / (t' - t) \\  (\forall (\rho, \lambda, \rho') \in \, \goes{}_S \, h', \dot{h} \not \models L(\rho')) \end{array}}
{(t,\rho,\mathit{false},h) \goes{(t',h',\epsilon)} (t',\rho,\mathit{false},h')} \\ 
\hline
\mathrm{(StopT)} \, \infr{\begin{array}{c} t < t' \quad \dot{h} = (h' - h) / (t' - t) \\ (\rho, \lambda, \rho') \in \, \goes{}_S \quad h', \dot{h} \models L(\rho') \end{array}}
{(t,\rho,\mathit{false},h) \goes{(t',h',\lambda)} (t',\rho',\mathit{true},h')} \\
\hline
\end{array}
$$
\caption{Rules defining the transition relation of a PELTS associated to a PEA.}
\label{table:rules-PELTS}
	\end{table}

Rule $\mathrm{(Steady)}$ models a step in which the PEA is in a steady state $\rho$ and remains in it because the equilibrium condition is still satisfied. The boolean flag of the PELTS state equal to \textit{true} indicates that the PEA is currently not adapting, i.e., it is in a steady state. The last observation of the PE value was $h$ at time $t$ (both values are in the current PELTS state). The new value of the PE $h'$ observed at the current time $t'$ is modelled as a transition of the PELTS labelled $(t',h',\epsilon)$, where $\epsilon$ indicates that the current PELTS transition is unnamed. The main pre-condition to apply the rule is that the current value of PE $h'$ and the current value of PE first derivative $\dot{h}$, computed as the difference quotient $(h' - h) / (t' - t)$, satisfy the equilibrium condition associated to $\rho$ (written as $h', \dot{h} \models L(\rho)$). Rule $\mathrm{(StartT)}$ models a step in which the PEA starts a transition because the equilibrium condition of the current steady state is no longer satisfied ($h', \dot{h} \not \models L(\rho)$). It is controlled that there exists on the PEA at least one outgoing transition from state $\rho$ and if this is so, the flag of the PELTS state changes from \textit{true} to \textit{false}. Rule $\mathrm{(ContT)}$ models a step in which the PEA is currently performing a transition and still remains in this situation because no one of the possible states that are reachable from the last steady state $\rho$ can be satisfied. This is expressed by the condition $\forall (\rho, \lambda, \rho') \in \, \goes{}_S \, h', \dot{h} \not \models L(\rho')$. Finally, rule $\mathrm{(StopT)}$ represents the end of a transition. The equilibrium condition of one of the possible PEA steady states reachable in one step from the last steady state $\rho$ (namely, one $\rho'$ such that $(\rho,\lambda, \rho') \in \, \goes{}_S$) is satisfied, thus the PEA ends the current transition. In the new PELTS state the steady state becomes $\rho'$ and the flag is put to \textit{true}. Note that the PELTS transition is labelled with the  name $\lambda$ that labels the PEA transition. Note also that only this kind of step is a source of non-determinism in the possible behavior of the PELTS. Indeed, if there are at least two (or more) possible PEA steady states, say $\rho'$ and $\rho'$, such that $(\rho, \lambda, \rho') \in \, \goes{}_S \, \wedge \,  h', \dot{h} \models L(\rho')$ and $(\rho, \lambda, \rho'') \in \, \goes{}_S \, \wedge \,  h', \dot{h} \models L(\rho'')$ then the PELTS could end up in state $(t',\rho',\mathit{true},h')$ or in state $(t',\rho'',\mathit{true},h')$ non-deterministically. 

From the definition above, it follows that a PELTS can execute paths of the form
$$
\pi = (0, \rho_0, \mathit{true}, 0) \goes{(t_1,h_1,\alpha_1)} (t_1,\rho_1,b_1,h_1) \goes{(t_2,h_2,\alpha_2)} (t_2,\rho_2,b_2,h_2) \cdots
$$
where $\alpha_i \in \Lambda \cup \{\epsilon\}$. 

Note that PETs, according to Definition~\ref{def:PEAObs}, can be naturally derivable from a path in a PELTS.

\begin{definition}[PETs from PELTS paths] \ \\
\label{def:traces-PEA-form-PELTS}
Consider a PEA $(R, \Lambda, \rho_0, \goes{}_S, L)$ and its associated PELTS. Each path of the PELTS
$$
\pi = (0, \rho_0, \mathit{true}, 0) \goes{(t_1,h_1,\alpha_1)} (t_1,\rho_1,b_1,h_1) \goes{(t_2,h_2,\alpha_2)} (t_2,\rho_2,b_2,h_2) \cdots
$$
corresponds to the following PET 
$$
tr = (t_1,h_1) \rightarrow (t_2,h_2) \rightarrow \cdots
$$ 
\end{definition}
	
On the other hand, a given PET $tr$ \textit{induces} a set of paths in the PELTS associated to the PEA. This is due to possible non-determinism among transitions in the PEA, as discussed above.

\begin{definition}[Form of PELTS paths induced from a PET]\ \\
\label{def:paths-PELTS-from-trace}
Consider a PEA $(R, \Lambda, \rho_0, \goes{}_S, L)$ and its associated PELTS. Let $tr$ be a PET. Then, putting $t_0 = 0$ and $h_0 = 0$, the trace $tr = (t_1,h_1) \rightarrow (t_2,h_2) \rightarrow \cdots$ induces a set of paths in the PELTS, each of which being of one of the following forms:

	\begin{enumerate}

\item \label{item:alternating} alternating forever between steady states and transitions:
$$
\begin{array}{l}
q_0 =(0, \rho_0, \mathit{true}, 0) 
   \, (\goes{(\cdot,\cdot,\epsilon)})^+ \\ 
(t_{j_0}, \rho_0, \mathit{false}, h_{j_0}) 
   \, (\goes{(\cdot,\cdot,\epsilon)})^* \\ 
(t_{j_0 + m_0}, \rho_0, \mathit{false}, h_{j_0 + m_0})   
   \goes{(t_{j_0 + m_0},h_{j_0 + m_0}, \lambda_0)} \\ 
(t_{j_1}, \rho_1, \mathit{true}, h_{j_1}) 
   \, (\goes{(\cdot,\cdot,\epsilon)})^+ \\
   \cdots \\
(t_{j_k + m_k}, \rho_k, \mathit{false}, h_{j_k + m_k}) 
   \, \goes{(t_{j_k + m_k},h_{j_k + m_k},\lambda_k)} \\
(t_{j_{k+1}}, \rho_{k+1}, \mathit{true}, h_{j_{k+1}})
   \, (\goes{(\cdot,\cdot,\epsilon)})^+ \\
   \cdots 
\end{array}
$$

\item \label{item:steadyforever} after a finite number of alternations, eventually reaching a steady state in which it remains forever:

$$
\begin{array}{l}
q_0 =(0, \rho_0, \mathit{true}, 0) 
   \, (\goes{(\cdot,\cdot,\epsilon)})^+ \\ 
(t_{j_0}, \rho_0, \mathit{false}, h_{j_0}) 
   \, (\goes{(\cdot,\cdot,\epsilon)})^* \\ 
(t_{j_0 + m_0}, \rho_0, \mathit{false}, h_{j_0 + m_0})   
   \goes{(t_{j_0 + m_0},h_{j_0 + m_0}, \lambda_0)} \\ 
(t_{j_1}, \rho_1, \mathit{true}, h_{j_1}) 
   \, (\goes{(\cdot,\cdot,\epsilon)})^+ \\
   \cdots \\
(t_{j_k + m_k}, \rho_k, \mathit{false}, h_{j_k + m_k}) 
   \, \goes{(t_{j_k + m_k},h_{j_k + m_k},\lambda_k)} \\
(t_{j_{k+1}}, \rho_{k+1}, \mathit{true}, h_{j_{k+1}})
   \, (\goes{(\cdot,\cdot,\epsilon)})^\infty 
\end{array}
$$

\item \label{item:transitionforever} after a finite number of alternations, eventually reaching a transition that is never ended:

$$
\begin{array}{l}
q_0 =(0, \rho_0, \mathit{true}, 0) 
   \, (\goes{(\cdot,\cdot,\epsilon)})^+ \\ 
(t_{j_0}, \rho_0, \mathit{false}, h_{j_0}) 
   \, (\goes{(\cdot,\cdot,\epsilon)})^* \\ 
(t_{j_0 + m_0}, \rho_0, \mathit{false}, h_{j_0 + m_0})   
   \goes{(t_{j_0 + m_0},h_{j_0 + m_0}, \lambda_0)} \\ 
(t_{j_1}, \rho_1, \mathit{true}, h_{j_1}) 
   \, (\goes{(\cdot,\cdot,\epsilon)})^+ \\
   \cdots \\
(t_{j_k + m_k}, \rho_k, \mathit{false}, h_{j_k + m_k}) 
    \, (\goes{(\cdot,\cdot,\epsilon)})^\infty
\end{array}
$$

\item \label{item:finite} if the PET is finite, the path is finite and may end in a steady state, i.e., the flag of the PELTS state is \textit{true}, or in the middle of a transition, i.e., the flag is \textit{false}.

	\end{enumerate}
\end{definition}

\subsection{Data-driven construction of the PEA for the \idion}
\label{sec:PEAIdion}

%%%%%%%%%%%%%%%%%%
\begin{figure}
\begin{center}
\includegraphics[width=\textwidth]{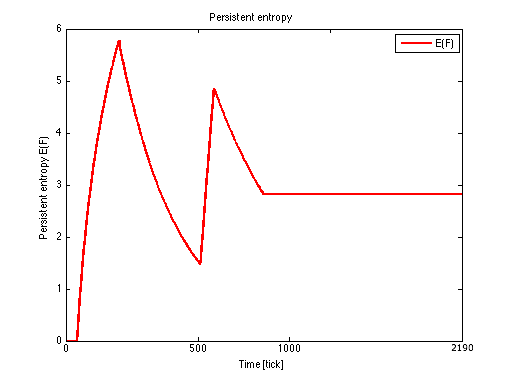}
\end{center}
\caption[PE]{PE of IS computed from a simulation of the \idion. For visualization purposes the plot has been smoothed but the local maxima and the plateaus have been preserved. The difference between the peaks amplitude is motivated by the fact that before the second peaks the antibodies have been already stimulated and the \textit{immune memory} has been reached, so the system is more reactive and is faster in the suppression of the antigen. The entropy value of the steady state is $H=2.87$.}
\label{fig:pe}
\end{figure}
%%%%%%%%%%%%%%%%%%

%%%%%%%%%%%%%%%%%%
\begin{figure}
\begin{center}
\includegraphics[width=\textwidth]{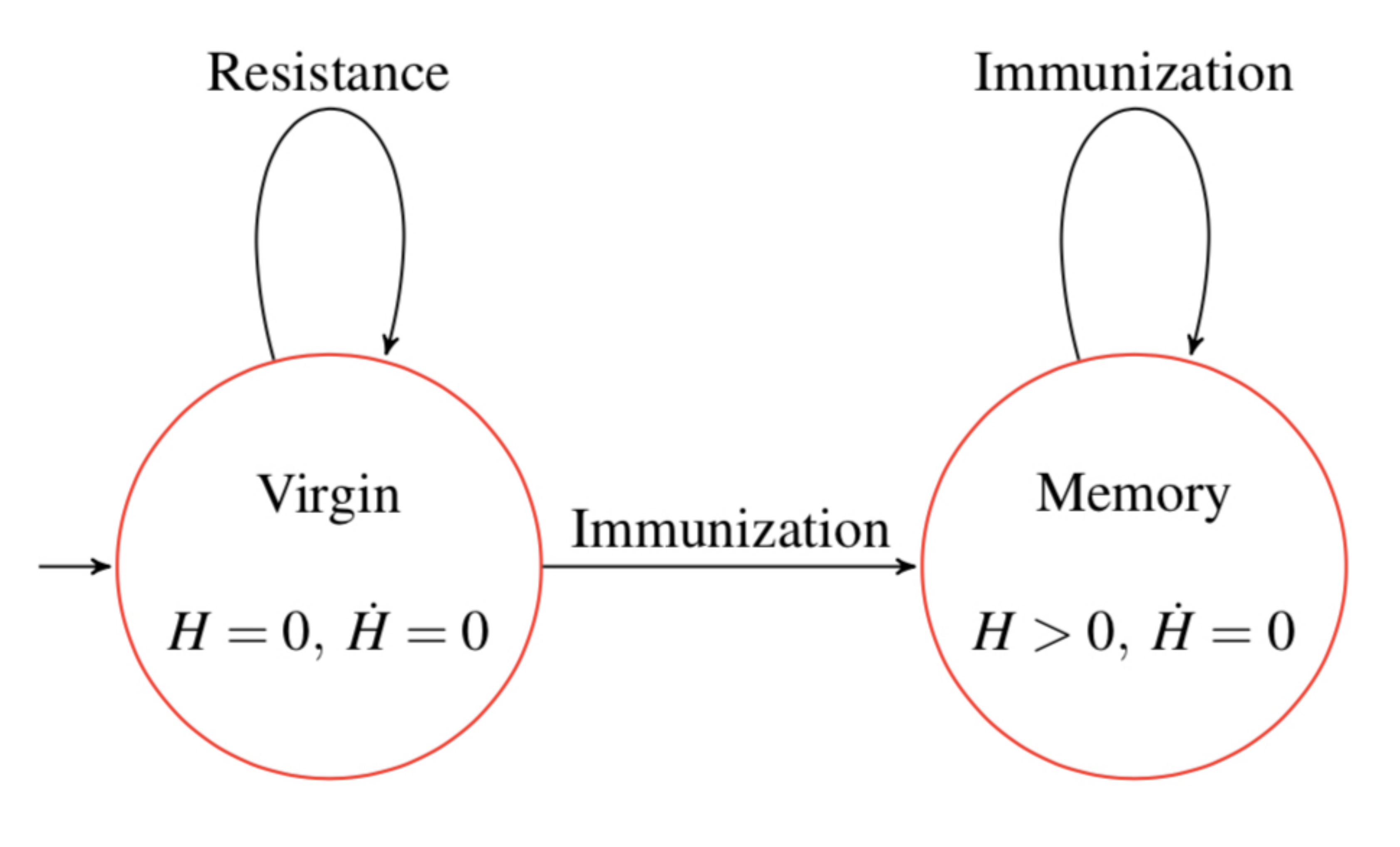}
\end{center}
\caption[Persistent entropy automaton]{PEA derived from simulated data for the \idion \negthickspace .}
\label{fig:SB1}
\end{figure} 
%%%%%%%%%%%%%%%%%%

In this section we recall how to derive the PEA from the simulation of \idion via \textit{C-ImmSim}. In \textit{C-ImmSim} each idiotype (both antigens and antibodies) is represented with a bit-string, in our case of 12-bit length. Two idiotypes $A_i$ and $A_j$ interact if and only if they are affine, that is their Hamming distance $d(A_i, A_j)$ is such that $11\leq d(A_i, A_j) \leq 12$. The pair-wise distances among all the idiotypes are stored in an \textit{affinity matrix}. However, the affinity matrix does not take into account the volume of the idiotypes that change as consequence of the volume of the antigens. In this case study, we decided to replace the affinity matrix with a \textit{coexistence matrix} $C$ where each element is a coexistent index. 
Given the Hamming distance $d(Ab_i(t),Ab_j(t))$ between two antibodies and their volumes $[Ab_{i}(t)], [Ab_{j}(t)]$ at time $t$, their 
\textit{coexistence index} is defined as follows:
\begin{equation}
C_{Ab_{i,j}}(t)=\frac{d(Ab_i(t),Ab_j(t)) \cdot [Ab_{i}(t)]\cdot[Ab_{j}(t)]}{\sum_{l=1}^{n} [Ab_{l}(t)]} 
\label{eq:eq1}
\end{equation}
where $n$ is the number of antibodies.

Equation~\ref{eq:eq1} expresses the fact that for lower values of affinity the volume must be more significant because the match between antibodies is less probable.

The coexistence matrix $C$ is a symmetric matrix and each element is taken to represent a weighted arc of an undirected graph. In this way we obtained a graph representation of our data taken from the simulation. 

We executed several (in the order of hundreds) simulations, each of them characterized by:
\begin{itemize}
\item a lifespan of 2190 discrete time ticks, where a tick corresponds to three days;
\item a repertoire of at most $10^{12}$ antibodies, i.e.\ the maximum number of antibodies available during the whole simulation;
\item an antigen volume $V = 10 \mu L$. 
\end{itemize}
Moreover, in order to increase the simulated real conditions we customized \textit{C-ImmSim} by adding the possibility to inject the antigen twice. The first injection is performed after a few ticks from the beginning of the simulation and the second injection is performed after an unknown (random) period.

The persistent homology of the filtered simplicial complex obtained from the weighted graph was computed with jHoles~\cite{binchi2014jholes, petri2013topological}. An example of the output is given in Table~\ref{tab:generators}. For each data sample, we obtained a number of $1$-dimensional holes in the order of hundreds. We did not obtain any persistent $n$-dimensional hole with $n>1$. 
%%%%%%%%%%%%%%%%%%
\begin{table}[ht]
\begin{center}
$$
\begin{array}{|l|ll|}
\hline
\beta_0 & [0.0 ; \infty ):& \{ [16] \} \\
\hline
 & [7.0 ; \infty): & \{ [ 320 , 3775 ] , [ 256 , 3775 ] , [ 320 , 3839 ] , [ 256 , 3839 ] \} \\
 \beta_1       & [6.0 ; \infty): & \{ [ 256 , 3839 ] , [ 256 , 3711 ] , [ 384 , 3711 ] , [ 384 , 3839 ] \} \\
        & [8.0 ; \infty): & \{ [ 260 , 3835 ] , [ 260 , 3839 ] , [ 256 , 3835 ] , [ 256 , 3839 ] \} \\
\hline
\end{array}
$$
\end{center}
\caption{Example of output of jHoles with \idion simulation data as input. One connected component appeared at filtration value 0.0, which is persistent. The generator is the vertex called 16. Three $1$-dimensional holes appeared at filtration values 7.0, 6.0 and 8.0, which are persistent. The four edges generating them are reported after the intervals.}
\label{tab:generators}
\end{table}
%%%%%%%%%%%%%%%%%
Figure~\ref{fig:pe} reports the chart of the average sequence of values of PE that were computed from the 2190 persistent barcodes for the considered simulation in the \idion case study  

As fully described in \cite{merelli2015topological}, we were able to recognize two steady states, namely \textit{virgin} and \textit{memory}. By analyzing the chart it is evident that two peaks are present. This reflects the fact that the system was stimulated twice and, thus, entered a critical transition followed by an adaptation phase from state \textit{memory} to itself. This then suggests that a self-transition should be added to the state \textit{memory} in the PEA.

The \textit{virgin} state of the \idion is characterized by the equilibrium condition $H=0 \, \wedge \,\dot{H} = 0$. This is the initial state in the PEA that we derived. The \textit{memory} steady state corresponds to a plateau in the chart and it is characterized by the equilibrium condition $H>0 \, \wedge \, \dot{H} = 0$. 
Finally, even if we did not observe it in our simulation, we know from domain specific knowledge that when in \textit{virgin} state, if the received stimulus is not grater than a certain threshold, then the immune activation does not start at all and after a while the system goes back to the initial state again. This means that also in the initial state of the PEA a self-loop transition should be added.
The derived PEA is depicted in Figure~\ref{fig:SB1}.

\section{Run-time Monitoring}
\label{sec:monitor}

In order to perform a data-driven topological run-time monitoring of a complex system,  we put all the already introduced pieces together. We assume that TDA was applied to a dataset produced by the complex system under study and that one or more sequences of PE observations, namely PETs, were obtained from the dataset. Moreover, we assume that a PEA was derived from the observation of the PE plot(s). Now, the objective is to define a framework for monitoring the behaviour of the complex system w.r.t.\ a property $\psi$. In particular, we want to study whether $\psi$ is satisfied or not given a certain finite behaviour of the complex system represented by a particular finite PET $tr$. 

We start by showing that the PEA can be used both as a model of the dynamics of the whole system and as a structure on which certain properties can be run-time monitored.
 
\begin{definition}[Monitor PEA] \ \\
A \emph{Monitor PEA} (MPEA) is a tuple $(R, \Lambda, \rho_0, \goes{}_S, L, AP, \Pi)$ where:
\begin{itemize}
\item $(R, \Lambda, \rho_0, \goes{}_S, L)$ is a PEA;
\item $AP$ is a set of boolean Atomic Propositions;
\item $\Pi \colon R \rightarrow 2^{AP}$ is a function labeling each steady state $\rho$ of the PEA with the atomic propositions that are true in $\rho$. 
\end{itemize} 
\end{definition}

For run-time verification purposes an MPEA can be used to ``execute'' the PEA against a given finite PET to obtain a trace, as follows. 
\begin{definition}[MPEA Executions and Traces] \ \\
Let $(R, \Lambda, \rho_0, \goes{}_S, L, AP, \Pi)$ be an MPEA and let $tr = (t_1,h_1) \rightarrow (t_2,h_2) \rightarrow \cdots \rightarrow (t_n,h_n)$ be a finite PET.
\begin{itemize}
\item An \emph{MPEA Execution} $e$ is \emph{one} of the sequences of steps of the PELTS associated to the PEA $(R, \Lambda, \rho_0, \goes{}_S, L)$ induced by the finite PET $tr$ according to Definition~\ref{def:paths-PELTS-from-trace}(\ref{item:finite}):
$$
e = (0, \rho_0, b_0=\mathit{true}, 0) \goes{(t_1,h_1,\alpha_1)} (t_1,\rho_1,b_1,h_1) \cdots  \goes{(t_n,h_n,\alpha_n)} (t_n,\rho_n,b_n,h_n)
$$
where $\alpha_i \in \Lambda \cup \{\epsilon\}$.
\item The \emph{MPEA Trace} $\sigma_e$, derived from $e$, is the sequence of sets of atomic propositions $\sigma_e = A_0 A_1 \cdots A_n$ such that:
$$
A_i = \left \{
\begin{array}{ll}
\Pi(\rho_i) & \mbox{if } b_i = \mathit{true}\\
\{\omega\} & \mbox{otherwise}\\
\end{array}
\right .
$$
\end{itemize}
\end{definition}

Note that, given a (finite) PET, the PELTS associated to a PEA can exhibit a set of (finite) executions due to non-determinism. In an MPEA execution we assume that such non-determinism has been resolved in order to obtain only one finite execution. Note also that the special atomic proposition $\omega$ labels all and only the states along the MPEA execution that correspond to the non-instantaneous firing of a PEA transition. All the other states of the MPEA execution are labelled with a subset of $AP$ according to the labeling function $\Pi$. 

MPEA Traces can be used to perform off-line run-time verification against formulas of an appropriate logic of interest. Several examples of suitable logics are described in \cite{leucker2009brief,bartocci2018introduction}.

\subsection{MPEA for the \idion}

An MPEA for the \idion case study can be defined by fixing the set of atomic propositions $AP = \{\mathit{virgin}, \mathit{memory}\}$ with the natural labeling function $\Pi(\mathrm{Virgin}) = \{\mathit{virgin}\}$ and $\Pi(\mathrm{Memory}) = \{\mathit{memory}\}$. The resulting MPEA is the one obtained from the PEA defined in Section~\ref{sec:PEAIdion} and these two new elements.

We considered a set of simulations of the \idion with \textit{C-ImmSim} to perform a first statistical analysis of the behaviours of the simulated system. Such analysis was particularly useful for deciding the minimum number of antibodies needed for properly reacting against the antigen or for establishing the right period of follow-up after the stimulation of the system.

We performed 1000 simulations and for each simulation we computed a PET by applying TDA to the produced data. The PET was executed on the MPEA resulting into an MPEA Execution and its corresponding MPEA Trace. We could classify the obtained traces into three groups, described in the following.

\begin{itemize}
\item[$I$] 198 traces terminated with $\{\mathit{virgin}\}$, of which:
	\begin{itemize}
		\item[a] 139 traces of the form $\{\mathit{virgin}\}^+$;
		\item[b] 59 traces of the form $\{\mathit{virgin}\}^+ \{\omega\}^+ \{\mathit{virgin}\}^+$;
	\end{itemize}
\item[$II$] 780 traces terminated with $\{\mathit{memory}\}$, of which: 
	\begin{itemize}
		\item[a] 429 traces of the form $\{\mathit{virgin}\}^+ (\{\omega\}^+ \{\mathit{memory}\}^+)^+$;
		\item[b] 351 traces of the form $\{\mathit{virgin}\}^+ \{\omega\}^+ \{\mathit{memory}\}^+$;
	\end{itemize}
\item[$III$] 22 traces terminated with $\{\omega\}$, of which: 
	\begin{itemize}
		\item[a] 18 traces of the form $\{\mathit{virgin}\}^+ (\{\omega\}^+ \{\mathit{memory}\}^+)^+ \{\omega\}^+$;
		\item[b] 4 traces of the form $\{\mathit{virgin}\}^+ \{\omega\}^+$.
	\end{itemize}
\end{itemize}

The traces in group $I.b$ correspond to the biological condition in which the antibodies space was completely destroyed by the antigens and so it appears empty (virgin state). The traces in group $II.a$ belong to systems that were stimulated multiple times. Traces in group $III$ correspond to system violation. Generally, these traces correspond to systems in which the repertoire of antibodies was not enough to eliminate the antigens. More in detail, traces $III.a$ regard systems stimulated multiple times but such that after certain stimulations the repertoire of antibodies was not sufficient for fighting against the antigen. Instead, traces $III.b$ belong to too short simulations.

To give some examples of properties that could be run-time verified on the case study of the \idion we consider the \textit{bounded} Linear Time Logic (bounded LTL). For a full description of this logic we refer to \cite{Latvala2004}.
The following are possible properties:
\begin{enumerate}
\item $\Box_{\leq 30} \neg \, \mathit{memory}$, informally meaning that ``the first immunization is never reached in the first 90 days'';

\item $\Box_{\leq 2190} \, (((\mathit{virgin} \vee \mathit{memory}) \wedge \bigcirc  \omega)) \Rightarrow \Diamond_{\leq 180} \; \mathit{memory})$, informally meaning that ``whenever an immunization phase is started, it ends within 1 year and a half'';

\item $\Diamond_{\leq 50} \Box_{\leq 150} \; \omega$ ``an immunization phase started, but the system was not able to reach the first immunization''.  
\end{enumerate}

Recall that each time tick of the simulation corresponds to 3 days, thus the mentioned time periods follows. Consider also that all the simulations we considered were long 2190 ticks, thus that is considered the limit of the bounded box.

\section{Conclusions}

In this paper we have formalized a new data-driven  run-time monitoring system for complex self-adaptive systems, the so-called Monitor for Persistent Entropy Automaton (MPEA). MPEA is obtained by leveraging a previous paper in which we have defined a data driven methodology for mining an automaton capable of mimicking the dynamics of a complex system, the so-called Persistent Entropy Automaton (PEA). PEA is obtained from time series of the persistent entropy that is a topological inviariant. The application of MPEA on the \idion has permitted to  demonstrated numerically that MPEA is suitable for classifying the behavior of complex self-adaptive system. We consider this work as the starting point for new interesting research directions. The methodology can be fruitfully used for discovering new behavioral patterns. These patterns allow to extract new knowledge that otherwise can not be captured or formally modelled. We also support that the methodology can be used for modeling complex software systems with the features of true concurrency self-adaptiveness.  Instead of the \idion, one could produce a graph representation of the system, e.g., a simplicial complex can be considered as the tuple formed by the computational resource and the processes that are acting against the resource. The application of the PEA can be used for modeling such systems without adopting the interleaving representation. The methodology can reveal if the system is performing a phase transition, e.g. when the system is acting against potential faults. Moreover, the MPEA can be used for formally verifying the execution of the system but regardless the order of execution of the actions, this preserves the simultaneous truly concurrent executions of the processes within a complex software system. 
In the future, we will work on the definition of a new run-time monitoring system for the controlling artificial neural network while used in safety critical systems, e.g. Advanced driver-assistance systems (ADAS) system, Unmanned aerial vehicle (UAV), etc\dots. Formal verification can be applied to establish functional correctness, but its scalability is limited due to the sheer complexity of these systems. To manage high complexity and limited specification resources, one alternative is to apply run-time monitoring techniques to detect when the system transitions into an unsafe state (i.e., one where it violates a critical safety requirement). However, due to the lack of exhaustive specifications of the environment in which autonomous systems are used the classical run-time monitoring techniques cannot be adopted and they must be enriched with new data-driven run-time techniques, similarly to PEA. The development and application of the new run-time monitoring system will represent an initial step to utilize runtime monitoring to achieve high assurance in the design of autonomous intelligent system.

\section{Acknowledgement}
M.R. would acknowledge Simone Fulvio Rollini for fruitful conversations. E.M. and L.T. would acknowledge the financial support of the Future and Emerging Technologies (FET) programme within the Seventh Framework Programme (FP7) for Research of the European Commission, under the FP7 FET-Proactive Call 8 - DyMCS, Grant Agreement TOPDRIM, number FP7-ICT-318121.

\bibliographystyle{plain}      % mathematics and physical sciences
\bibliography{paperRV}   % name your BibTeX data base

%\begin{thebibliography}{}
%%
%% and use \bibitem to create references. Consult the Instructions
%% for authors for reference list style.
%\end{thebibliography}

\end{document}